\documentclass[twocolumn,preprintnumbers,amssymb,amsmath,9pt,superscriptaddress]{revtex4}[10pt]
\usepackage{graphicx,psfrag}
\usepackage{dcolumn}
\usepackage{color}
\usepackage{bm}
\begin{document} 
\input epsf.tex
\newcommand{\beq}{\begin{eqnarray}}
\newcommand{\eeq}{\end{eqnarray}}
\newcommand{\nn}{\nonumber}
\def\ltap{\ \raise.3ex\hbox{$<$\kern-.75em\lower1ex\hbox{$\sim$}}\ }
\def\gtap{\ \raise.3ex\hbox{$>$\kern-.75em\lower1ex\hbox{$\sim$}}\ }
\def\CO{{\cal O}}
\def\CL{{\cal L}}
\def\CM{{\cal M}}
\def\tr{{\rm\ Tr}}
\def\CO{{\cal O}}
\def\CL{{\cal L}}
\def\CM{{\cal M}}
\def\mpl{M_{\rm Pl}}
\newcommand{\bel}[1]{\be\label{#1}}
\def\al{\alpha}
\def\bt{\beta}
\def\eps{\epsilon}
\def\eg{{\it e.g.}}
\def\ie{{\it i.e.}}
\def\mn{{\mu\nu}}
\newcommand{\rep}[1]{{\bf #1}}
\def\be{\begin{equation}}
\def\ee{\end{equation}}
\def\bea{\begin{eqnarray}}
\def\eea{\end{eqnarray}}
\newcommand{\eref}[1]{(\ref{#1})}
\newcommand{\Eref}[1]{Eq.~(\ref{#1})}
\newcommand{\gsim}{ \mathop{}_{\textstyle \sim}^{\textstyle >} }
\newcommand{\lsim}{ \mathop{}_{\textstyle \sim}^{\textstyle <} }
\newcommand{\vev}[1]{ \left\langle {#1} \right\rangle }
\newcommand{\bra}[1]{ \langle {#1} | }
\newcommand{\ket}[1]{ | {#1} \rangle }
\newcommand{\fb}{\,{\rm fb}^{-1}}
\newcommand{\ev}{{\rm eV}}
\newcommand{\kev}{{\rm keV}}
\newcommand{\Mev}{{\rm MeV}}
\newcommand{\gev}{{\rm GeV}}
\newcommand{\tev}{{\rm TeV}}
\newcommand{\mev}{{\rm MeV}}
\newcommand{\meV}{{\rm meV}}
\newcommand{\mnu}{\ensuremath{m_\nu}}
\newcommand{\nnu}{\ensuremath{n_\nu}}
\newcommand{\mlr}{\ensuremath{m_{lr}}}
\newcommand{\acc}{\ensuremath{{\cal A}}}
\newcommand{\mav}{MaVaNs}
\newcommand{\disc}[1]{{\bf #1}} 
\newcommand{\mh}{{m_h}}
\newcommand{\hb}{{\cal \bar H}}
\def\R{{\mathbb R}}
\def\S{{\mathbb S}}
\def\Z{{\mathbb Z}}

\newcommand{\add}[1]{{#1}}

\title{Strings from domain walls in supersymmetric Yang-Mills theory and adjoint QCD}
\author{Mohamed M. Anber}
\affiliation{Institute de Th\' eorie des Ph\' enomen\` es Physiques, \' Ecole Polytechnique F\' ed\' erale de Lausanne, CH-1015 Lausanne, Switzerland }
\author{Erich Poppitz}
\affiliation{Department of Physics, University of Toronto, 
Toronto, ON M5S 1A7, Canada}
\author{Tin Sulejmanpa\v si\' c}
\affiliation{Department of Physics, North Carolina State University, Raleigh, NC 27695, USA }
\date{\today}
\begin{abstract} 
We study strings between static quarks in QCD with $n_f$ adjoint fermions, including $\mathcal N=1$ Super Yang-Mills (SYM), in the calculable regime on $\R^3\times \S^1$, \add{which shares many features with the XY-spin model.} We find that they have many qualitatively new features not previously known. The difference from other realizations of abelian confinement is  due to the composite nature of magnetic bions, whose Dirac quantum with fundamental quarks is two, and  to the unbroken part of the Weyl group.  In particular we show that strings are composed of two domain walls, that quarks are not confined on domain walls, that strings can end on domain walls, and   that  ``Y" or ``$\Delta"$ baryons can form. \add{By similar argumentation, liberation of vortices on domain walls in the condensed matter counterparts may have important implications in the physics of transport. In the gauge theory we briefly discuss the lightest modes of strings and the decompactification limit.}

\end{abstract}

\maketitle
  
While ubiquitous in nature, color confinement is one of the least-understood features of  Yang-Mills (YM) theory. Theoretically controlled approaches usually involve models that differ, in various ways, from real-world QCD. Nonetheless, one's hope is that their study will reveal features of confinement that transcend the particular model. 

A rare theoretical laboratory where confinement  is under theoretical control within field theory is offered by QCD(adj): an $SU(N_c)$ YM theory with a strong scale $\Lambda$ and $n_f$ Weyl fermions in the adjoint representation, compactified on $\R^{1,2} \times \S^1_L$ with fermions periodic  around the spatial $\S^1_L$ of size $L$. \" Unsal showed \cite{Unsal:2007jx} that  for $L N_c \Lambda$ $ \ll$ $ 1$   confinement 
is due to  the proliferation of topological molecules---the magnetic bions. These  are non-self-dual correlated tunneling events composed of various fundamental  and twisted \cite{Lee:1997vp,Kraan:1998pm} monopole-instantons. For small but finite $LN_c\Lambda$, magnetic bion confinement extends the 3d Polyakov mechanism of confinement  \cite{Polyakov:1976fu} to locally-4d theories qualitatively different from 3d theories with fermions where confinement  is lost \cite{Affleck:1982as}. In passing, we note that
QCD(adj), bions and their constituents are studied in connection with deconfinement, resurgence, theta-dependence and volume independence, e.g.~\cite{Anber:2011gn, Poppitz:2012sw, Argyres:2012ka,Basar:2013sza,Anber:2013sga,Shuryak:2013tka,Poppitz:2013zqa,Anber:2014sda,Bruckmann:2014sla,Bergner:2014dua,Misumi:2014raa,Misumi:2014bsa,Larsen:2014yya}.
 
The goal of this paper is to study confining strings in QCD(adj)/SYM in the calculable regime. We find that they have an interesting structure, to the best of our knowledge not previously discussed in  confining strings studies. We shall see that, while still abelian in nature, QCD(adj) strings retain more features expected in the nonabelian theory compared to other theories with abelian confinement. \add{In addition the entire discussion relies only on the effective  Lagrangians \eqref{su2string1} or \eqref{suNstring1} below. Since these have been related to the condensed matter systems \cite{Anber:2011gn,Anber:2012ig,Anber:2013doa,Anber:2014lba}, most of our observations may have direct important consequences in the physics of these systems. }

At distances $\gg$$L$ massless QCD(adj) on $\R^3 \times \S^1_L$ dynamically abelianizes \cite{Unsal:2007jx}.  Ignoring fermions for the moment, for $SU(2)$ gauge group the effective bosonic (Euclidean) Lagrangian is:
 \begin{eqnarray}
 \label{su2string1}
{\cal L}_B = &&  M \left[ (\partial_\mu \sigma)^2 + (\partial_\mu \phi)^2 + {m^2 \over 2} (\cosh 2 \phi - \cos 2 \sigma)  \right. \nonumber \\
&& ~~+ ~(n_f -1) V_{pert.} (\phi) \left.  \right]~.
 \end{eqnarray}
The scales and fields   in (\ref{su2string1}) are as follows. The scale   $M$ is of order $g^2/L$, where $g^2$ is the weak  4d gauge coupling at the  scale $1/L$. The   scale  $m \sim \exp({- {4\pi^2/g^2}}) M$ is nonperturbative (${4\pi^2/g^2}$  is  the action of a   monopole-instanton) and exponentially small. 
With exponential-only accuracy (for pre-exponential factors, see \cite{Anber:2011de,Argyres:2012ka}) one can  think of $M$ as the cutoff scale of our effective theory and of $m$ as the mass scale of  infrared physics. 

The long-distance theory (\ref{su2string1}) has two bosonic  fields. The field $\phi$ describes the deviation of the trace of the Wilson line around $\S^1_L$ from its center symmetric value. Equivalently, $\phi$ is the radial mode of the adjoint Higgs  (the Wilson line) breaking $SU(2) \rightarrow U(1)$. The field $\sigma$ is dual to the photon in the unbroken Cartan subalgebra (the   $\tau^{(3)}$ direction) of $SU(2)$. In Minkowski space  $M \partial_0 \sigma \sim F_{12}^{(3)}$ is the magnetic field, and $M \epsilon_{ij}\partial_{j}  \sigma \sim {E_{i}}^{(3)}$ is the electric field (where $\epsilon_{ij}=-\epsilon_{ji}$ and $i,j=1,2$). 

As is clear from the  discussion of scales, the terms  in (\ref{su2string1})  proportional only to $M$  are perturbative. We shall not need the explicit expression \cite{Unsal:2007jx,Argyres:2012ka} for the perturbative potential $V_{pert}(\phi)$. This term is absent for  SYM ($n_f$$=$$1$). For $n_f$$>$$1$, $V_{pert}$ stabilizes $\phi$ at the center symmetric value and  gives it mass of order $M$, hence $\phi$ can be integrated out. The relative normalization between the  potentials for $\phi$ and $\sigma$  given in (\ref{su2string1}) is for SYM. 

Of most interest to us is the origin of the nonperturbative terms in (\ref{su2string1}). The nonperturbative  potential for $\phi$, $\sim$$\cosh 2 \phi$, is due to neutral bions \cite{Poppitz:2011wy, Argyres:2012ka} and will not play an important role here (except for being the only source stabilizing $\phi$ at the center-symmetric value $\phi$$=$$0$ in  SYM). The other term, $\sim$$\cos 2 \sigma$, is of utmost importance to us, as  it captures the effect of the magnetic bions---the leading cause of confinement in QCD(adj). The factor of two in the argument of the cosine reflects their composite nature: they have magnetic charge two while fundamental monopole-instantons have unit charge.
This term is responsible for the generation of  mass gap for gauge fluctuations (mass $m$ for the dual photon $\sigma$) and for the confinement of electric charges. The theory (\ref{su2string1}) has two vacua $\sigma$$=$$0, \pi$, both with $\phi$$=$$0$, corresponding to the spontaneous  breaking of the anomaly-free discrete chiral symmetry (the $R$-symmetry in  SYM).

 Confinement is detected by   the area law for the Wilson loop in a representation $\cal R$, taken along a closed contour $C$,  $W(C,{\cal R})$$\equiv$$\tr_{{\cal R}} P \exp({i \oint_C A})$. For an $SU(2)$  fundamental representation,  we need to compute the expectation value of  $W(C, {1\over 2}) $$\; \sim\; $$ \exp({ {i \over 2} \oint_C A^{(3)}}) $$ = $$\; \exp({{i \over 2} \int_S B^{(3)}})$. Here $A^{(3)}$ is the (electric) gauge field in the Cartan direction, $B^{(3)}$$=$$d A^{(3)}$ is its field strength, and $S$ is a   surface spanning $C$ (the omitted second contribution to the trace of the fundamental Wilson loop gives an identical area law). 
 
\begin{figure}
\begin{center}
\includegraphics[width=0.46\textwidth]{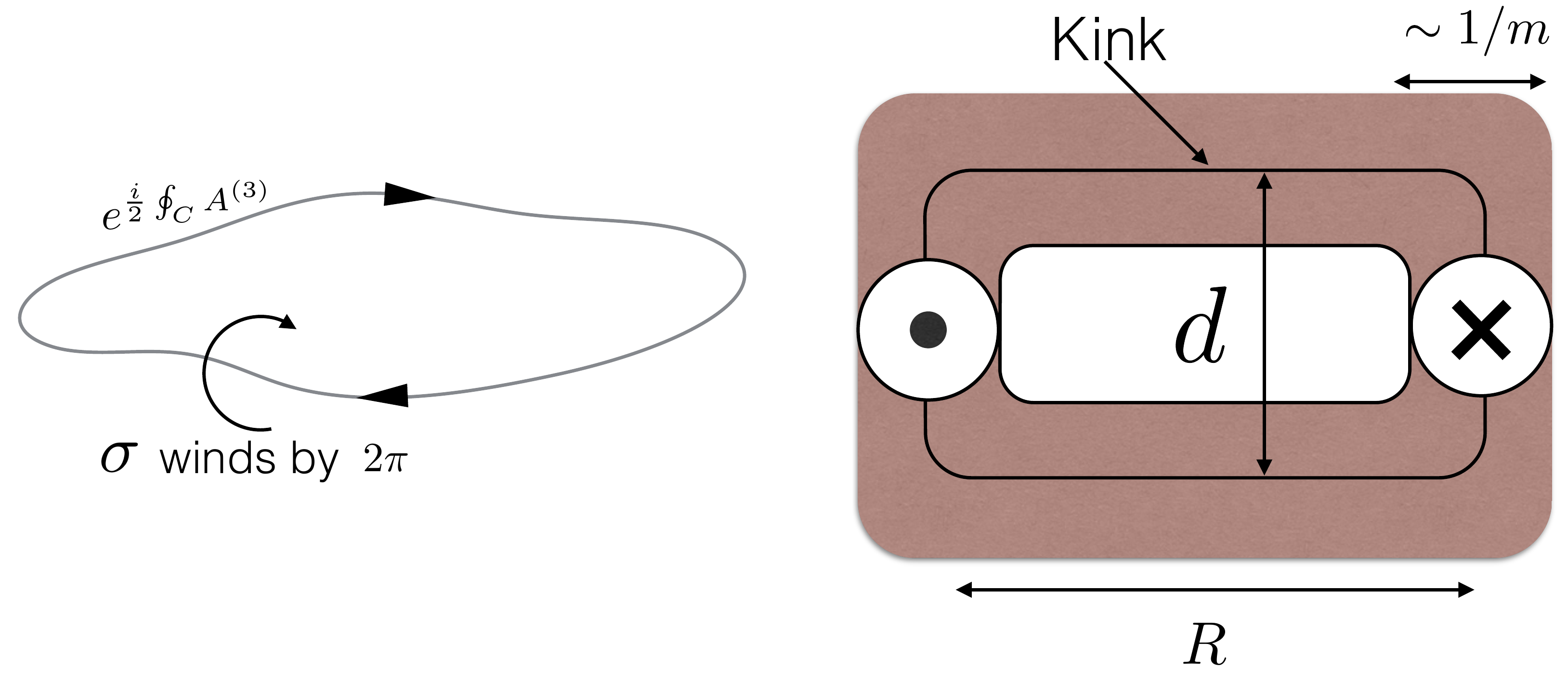}
\caption{Left: the Wilson loop and the monodromy of $\sigma$. Right: Sketch of the confining string configuration $\bar\sigma$ with the correct monodromy, composed of two domain walls. The dot and cross represent probe quarks a distance $R$ apart. The maximum distance between the walls, of    thickness  $1/m$, is $d$.  
} \label{fig:111}
\end{center}
\end{figure}

Insertion of the Wilson loop in the dual language of the $\sigma$ field (recalling that $\sigma$$\sim$$\sigma +2\pi$) amounts to the following instruction \cite{Polyakov:1976fu}: erase the contour from the space, and have $\sigma$ wind by $2\pi$ for any contour which has linking number one with the Wilson loop---a $2\pi$ monodromy (see left panel of Fig.~\ref{fig:111}).
Take  a  rectangular contour in the $y$$-$$x$-plane ($y$ is Euclidean time)  with span $T\; (R)$ in the $y\; (x)$-direction. For infinite $R$ and $T$,  $\sigma$ jumps by $2 \pi$ upon crossing the $y$$-$$x$ plane.
If the potential in (\ref{su2string1}) was---as in Polyakov's original 3d SU(2) gauge theory with an adjoint Higgs field---$\cos\sigma$, the  field configuration extremizing the action (\ref{su2string1}) with the correct monodromy, which we denote $\bar\sigma$, would be equivalent \cite{Polyakov:1976fu}  to a domain wall  with $y$$-$$x$-plane worldvolume, where $\bar\sigma$ would change by $2\pi$ as $z$ varies between $\pm\infty$. We would have $W(C, {1 \over 2}) \sim e^{- \Sigma_{str} R T}$, with string tension $\Sigma_{str}$  proportional to the domain wall tension (for a recent review see \cite{Anber:2013xfa}).

The physical difference between monopole-instanton confinement in the  3d Polyakov model  and QCD(adj) on $\R^3 \times \S^1$---the fact that the magnetic bions have magnetic charge two---is reflected in the  $\cos 2 \sigma$ potential  (\ref{su2string1}). Now, the $\bar\sigma$-field configuration  with the right monodromy has to be more complicated than a single domain wall. 
To study it, we keep the time ($y$) extent of $C$ infinite and consider a finite spatial ($x$) extent $R$. As the  $\bar\sigma$ configuration has monodromy $2 \pi$ across $C$, in this simple one-field case it is clear that  (since the periodicity of the $\cos 2 \sigma$ potential is $\pi$) the string has to be  composed of two domain walls. To get a picture of the extremal configuration, consider   Fig.~\ref{fig:111}, with  parameters $R,d$ defined in the caption. A sketch of a two-domain wall configuration is shown, with the second infinite worldvolume direction (the time $y$) perpendicular to the page.  The action has two parts, excluding contributions from the junctions (subleading at large $R$): the tension of the two domain walls, proportional to twice their area (we take $T (R +d)$ as the area) and the wall-wall long-distance repulsion ($\sim$$ e^{- m d}$). Thus,  $S \sim M m T (R + d)+M m T R e^{- m d}$, up to   numerical factors. Extremizing with respect to $d$, we find  $m d_* \sim \log { m R}$, a logarithmic growth of the transverse size of the confining string configuration with the separation between the probe charges.

\begin{figure}
\begin{center}\includegraphics[width=2.9in]{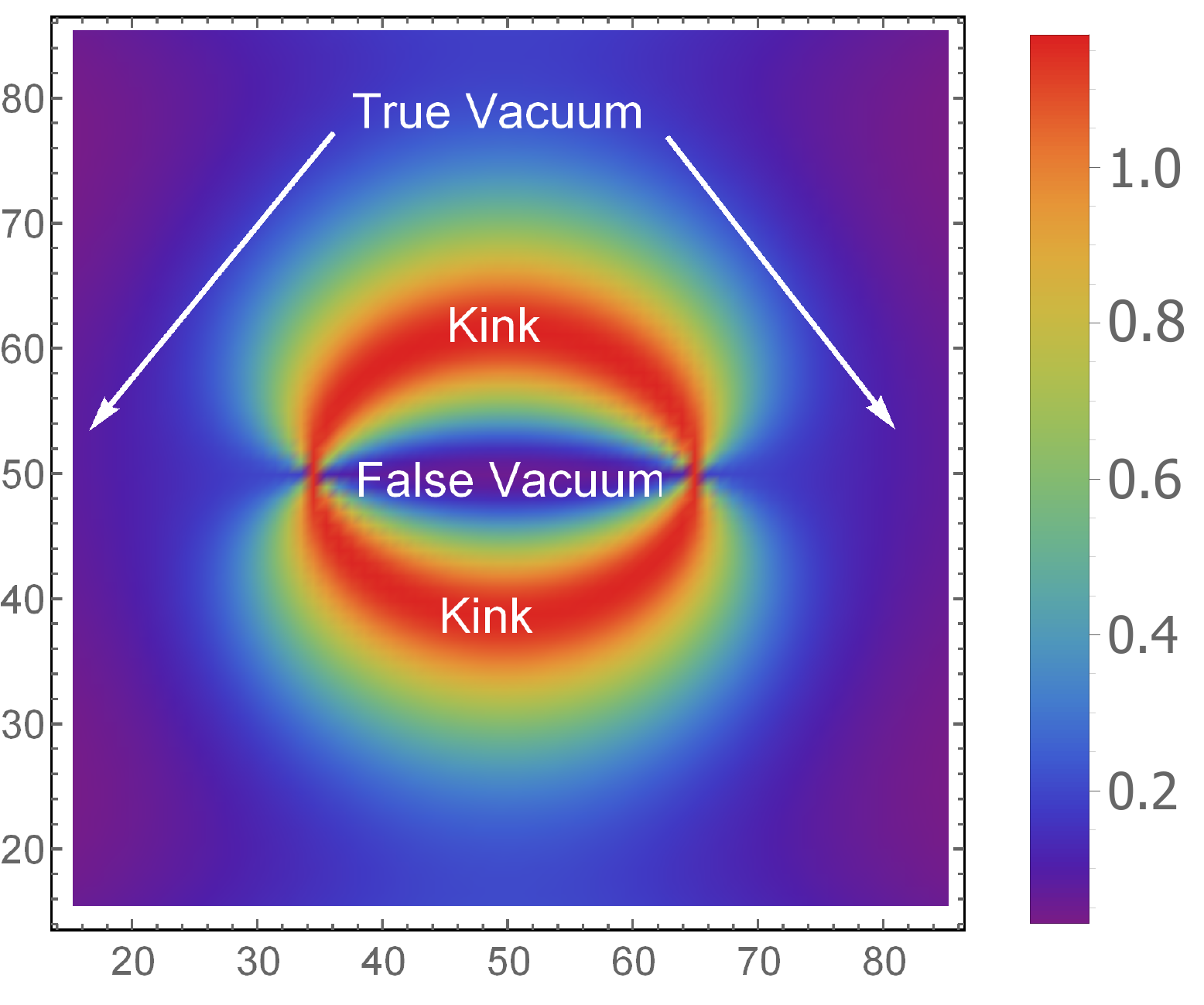} 
\caption{The action density of the confining string $\bar\sigma$ obtained by numerically minimizing, via Gauss-Seidel relaxation,  the action (\ref{su2string1}) with the correct monodromies. The lattice has spacing $1/M$, size $100 \times 100$, and $M/m=20$.  The classical $\log R$ growth of the transverse separation from  the model of Fig.~\ref{fig:111} is also seen to hold upon studying different size strings.} \label{fig:222}
\end{center}
\end{figure}

Remarkably, the above simple model  captures the behavior of the actual extremum of (\ref{su2string1}), shown on Fig. \ref{fig:222}, including the $\log R$ growth of the transverse size. Our remarks so far also hold for deformed-YM theory   \cite{Unsal:2008ch}, where, for $\theta$$=$$\pi$ \cite{Unsal:2012zj} the single monopole contribution vanishes.

The adjoint fermions were, so far, ignored. Their Cartan components have an effective  Lagrangian \cite{Unsal:2007jx}  \begin{eqnarray}
 \label{su2string1fermion}
{\cal L}_F = M \left[ i \bar\lambda \bar\sigma^\mu \partial_\mu \lambda +   {m \cos  \sigma \over 2 M^{n_f - 1}}   \left[ (\lambda  \lambda)^{n_f} + {\rm h.c.} \right] \right]~.
 \end{eqnarray}
We omitted, for brevity, a summation over the $n_f$ flavor indices in the kinetic term and a product over the flavor indices  in the interaction term (the 't Hooft determinant in the monopole-instanton background). The field $\phi$ is also set to its vanishing vev. For  SYM, apart from omitting $\phi$,  (\ref{su2string1fermion}) has correct normalization. It is, in fact,  the effect of the fermions on the 
confining string where the difference between SYM and QCD(adj) with $n_f$$>$$1$ shows up most profoundly. 

In SYM, the fermions are massive in the $\sigma$$=$$0, \pi$ vacua. They have exact zero modes in a single domain wall background, with exponential fall off away from the wall. Because of the gap $m$ in the bulk, the fermion induced wall-wall interaction  is expected to be exponentially suppressed, $\sim$$m^2 e^{- c m d}$, $c$$\ge$$1$ (a calculation of the determinant,  requiring some mild background modeling even for parallel walls, yields attraction with $c$$>$$1$). The fermion-induced exponential interaction at large $d$ is further accompanied by an ``$\hbar$"$\sim$${m\over M}$  loop suppression factor, hence the classical bosonic repulsion between the walls $\sim M m e^{- m d}$ dominates.  
 Thus, in SYM  the logarithmic growth of the transverse string size is not affected by the fermions.  
The $\log R$ growth of the string transverse size   is reminiscent of the behavior of magnetic strings (ANO vortices) which confine monopoles on the Higgs branch of $N$$=$$2$ SQCD \cite{Yung:1999du}. However,  the underlying semiclassical physics  is  different; in particular, in contrast to   \cite{Yung:1999du}, our strings in SYM obey the usual area law with tension $\sim$$M m$. 

In contrast to SYM, in non-supersymmetric QCD(adj) with $n_f$$>$$1$ the Cartan components of the $n_f$ Weyl adjoints are massless, due to the unbroken $SU(n_f)$ chiral symmetry. Thus, despite the fact that their interaction with the wall in (\ref{su2string1fermion}) is highly suppressed,   they induce a power-law force    competing  with the exponential repulsion  at large $d$. The leading effect of the fermions occurs at  $2 n_f$$-$$1$ loop order; its calculation, of which we just give the result,  is similar in spirit to Casimir energy calculations. Fermion loops are found to generate  a wall-wall attraction at large $d$. Per unit volume, it is $\sim-m^2 \left(m\over M \right)^{4 n_f} {(m d)^{-4 n_f + 4}}$, dominating  the bosonic repulsion $\sim M m e^{- m d}$ at large $d$. The expression for the action of our toy model, with fermion attraction included, is $S = T (R + d) M m + R T M m e^{- m d} - R T m^2 \left(m\over M \right)^{4 n_f}\slash (m d)^{4 n_f - 4}$. The extremum condition (to which the area term does not contribute for large $T$) is now  $e^{- md} \sim {e^{- {4 \pi^2} (4 n_f +1)\slash g^2} \slash (m d)^{4 n_f - 3}}$. At small $g^2$, we thus have $m d_* \approx {4 \pi^2 (4 n_f + 1) \slash g^2}$,  a stable wall-wall separation parametrically large compared to the single domain wall width. 
Numerical confirmation of the stabilized transverse size $d_*$ of the string is challenging, but our estimate of the size stabilization is reliable at small $g$ and large $R$. 

As a  consequence of the stabilized transverse size of the confining string in  $n_f$$>$$1$ QCD(adj), the second translational Goldstone mode, the ``breather" mode of the two walls, is now gapped even at infinite $R$. The gap for this mode, $m_{br}$, can be estimated by taking the second derivative of the wall-wall interaction potential at $d_*$, $m_{br} \sim m e^{- {4 \pi^2} 2 n_f\slash g^2 }$. The breather mode mass $m_{br}$ is   a new  scale on the string worldsheet,  well below the ``glueball"---the bulk mass gap  $m$  for gauge fluctuations. 
 
\begin{figure}[tbp] 
   \centering
   \includegraphics[width=.37\textwidth]{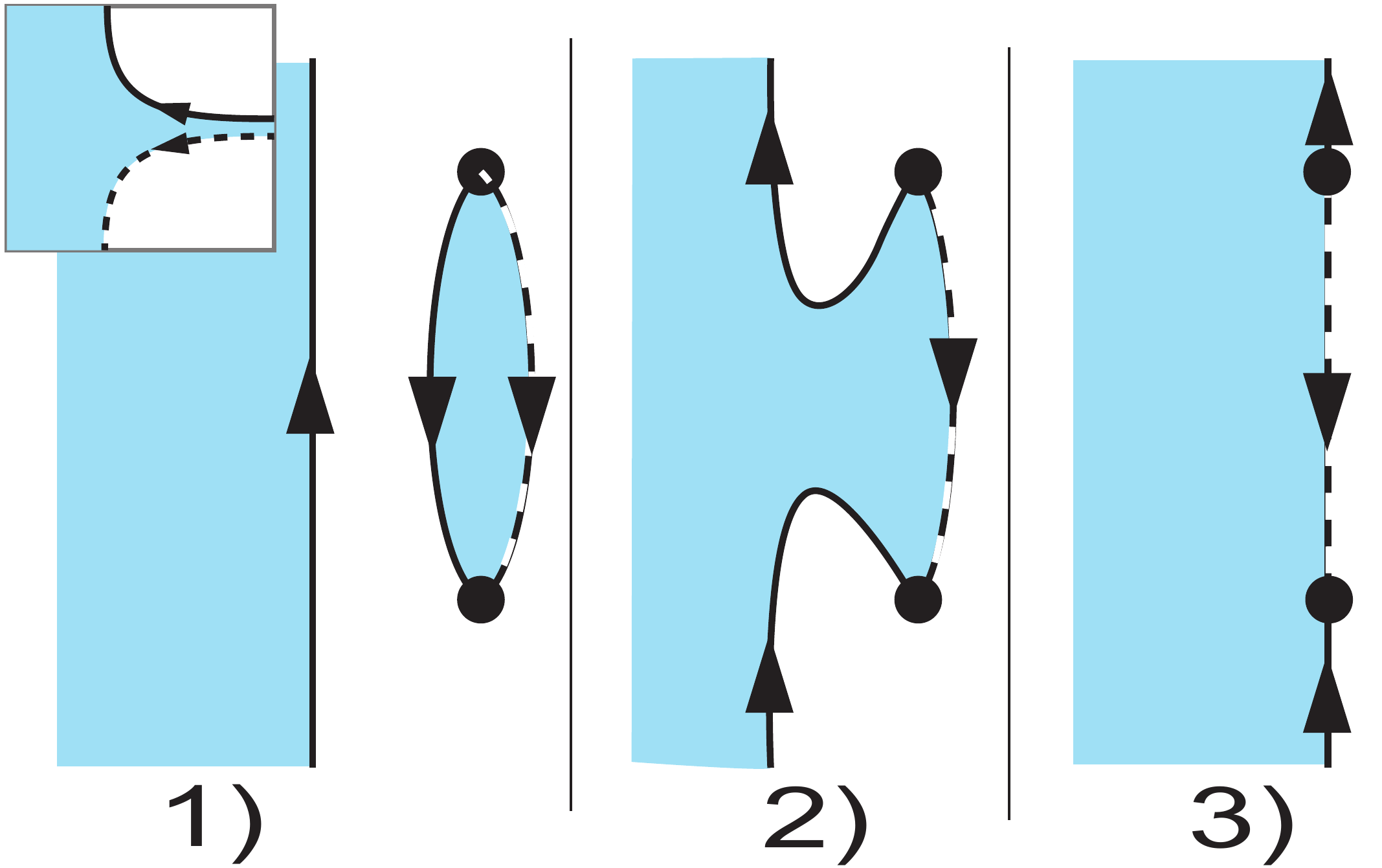} 
   \caption{A sketch of how a $q\bar q$ pair can fuse into the DW (from left to right). The shaded and white regions represent distinct vacua of the theory. The solid black line represents the BPS$_1$ DW, while the dashed line represents the anti-BPS$_2$ DW, while the arrows represent their electric fluxes. The black dots are the quark and the anti-quark. The inlay in the upper left corner shows  a fundamental string ending on a DW. }
   \label{fig:fuse}
\end{figure}

The fact that the strings are composed out of domain walls (DW) -- a situation opposite to what was suggested in \cite{Armoni:2003ji} -- has drastic implications on how the fundamental quarks interact with DWs. For $SU(2)$  there are two types of DWs, which we  label  BPS$_1$ and BPS$_2$, and their anti-walls. The distinction is in the electric fluxes which they carry, but they both satisfy the same BPS equation, e.g.~\cite{Hori:2000ck}. The fundamental string is made out of the BPS$_1$ and an anti-BPS$_2$, where each carries $1/2$ of the fundamental electric flux. If a quark anti-quark ($q\bar q$) pair is in the vicinity of the DW, however, the DW flux can cancel part of the flux of a $q\bar q$ pair, and absorb it into its worldsheet, see Fig. \ref{fig:fuse}. The $q\bar q$ pair on the DW would then be liberated, as all the tension of the pair has been absorbed into the DW tension. This leads to deconfinement in the DW worldsheet. This is reminiscent of the DW localization, where a theory in the DW worldsheet is in Coulomb phase, so that quarks are liberated \cite{Dvali:1996xe}. \add{We also note that in a certain Higgs vacuum of 4d theories, monopole--anti-monopole pairs have support on stable non-abelian (\emph{magnetic}) strings \cite{Shifman:2012yi,Shifman:2014qfa}. In this work the strings are genuine \emph{electric} strings.}

Deconfinement of quarks on the DW also implies that strings can end on DWs (see inlay of Fig.~\ref{fig:fuse}).
In MQCD, SYM strings have been argued to end  on DWs and a heuristic explanation  by S.-J. Rey \cite{soojongrey}, using the vacuum structure and ideas about confinement,  is given in \cite{Witten:1997ep}. The phenomenon was subsequently explored from modeling the effective actions of the Polyakov loop and gaugino condensates \cite{Campos:1998db}. Here, we   found---for the first time, to the best of our knowledge---an explicit realization of this phenomenon in a field theory setting where the confining dynamics is understood \footnote{That \emph{magnetic} strings can end on DWs has been known for some time, however \cite{Shifman:2002jm}.}. 

Our  discussion of confining strings in QCD(adj) generalizes to the higher-rank case. We shall  focus  only on a few salient points. All fields in (\ref{su2string1}) become  $N_c$$-$$1$ dimensional vectors, describing the light degrees of freedom left after  $SU(N_c)$$\rightarrow$$U(1)^{N_c -1}$ breaking.
It suffices to study  the operator
$W(C,\lambda) = e^{i \vec\lambda  \cdot \oint_C \vec{A}^{(3)} }$, with $\vec{\lambda}$---a weight of   $\cal R$ (a vector of $U(1)^{N_c-1}$ electric charges), as   the trace of the Wilson loop is obtained by summing over all weights of $\cal R$. As in (\ref{su2string1}), semiclassically $\langle W(C,\lambda) \rangle \sim e^{- S_{class}[\bar\sigma(C)]}$, with the magnetic bion potential
\begin{equation}
\label{suNstring1}
{\cal L}_{bion}=  -m^2 M\; \sum\limits_{i=1}^{N_c}\cos\left[ (\vec\alpha_i^* - \vec\alpha_{i+1 ({\rm mod}\; N_c)}^*) \cdot \vec\sigma  \right]\;,
\end{equation}
 replacing the one  in (\ref{su2string1}). 
Here $\vec{\alpha}_i^*$  label the  simple ($i$$<$$N_c$), affine ($i$$=$$N_c$) coroots ($|\vec{\alpha}_i^*|^2$$=$$2$); $M$ and $m$ are, up to irrelevant factors, as in (\ref{su2string1}).  The fields $\vec\phi$ are   set to their vev $\vec\phi$$=$$0$; the full Eq.~(\ref{suNstring1}) is in \cite{Argyres:2012ka} for $n_f$$>$$1$ and \cite{Anber:2014lba} for SYM (to get back (\ref{su2string1}), use $\alpha_1^*$$=$$- \alpha_2^*$$=$$\sqrt{2}$, $\lambda$$=$$1/\sqrt{2}$ and redefine $m,M,\sigma$). Clearly, 
a string between  quarks with charges $\vec{\lambda}$ should have $2 \pi \vec{\lambda}$ monodromy  of  $\vec\sigma$ around  $C$.

 An important fact, with crucial consequences for the string spectrum,  is that,  due to the existence of the  twisted (affine) monopole-instanton \cite{Lee:1997vp} and the preserved center symmetry, a $\Z_{N_c}$ subgroup of the Weyl group, cyclically permuting the $N_c$ roots in (\ref{suNstring1}), is unbroken in    QCD(adj). Denoting by  $P$  the generator of the  cyclic Weyl group, using an $N_c$-dimensional basis  for the roots (one  linear combination of the $N_c$ $\sigma_k$'s decouples \cite{Argyres:2012ka}),   its action is:  $P \sigma_k$$=$$\sigma_{k+1 ({\rm mod} N_c)}$, or    $P \vec{\alpha}_k$$=$$\vec\alpha_{k+1 ({\rm mod} N_c)}$. The $P$ symmetry ensures that  strings confining quarks in $\cal R$ of $SU(N_c)$ have equal tension for all weights of $\cal R$ that lie in the same orbit of the cyclic Weyl subgroup. Since  $P$  permutes the $N_c$ weights of the fundamental representation, strings confining any component of the fundamental quarks have equal tension. This is different from Seiberg-Witten theory where the Weyl group is completely broken \cite{Douglas:1995nw}. Still, the multiplicity of meson Regge trajectories in the calculable regime of QCD(adj) is different from that expected in the full nonabelian theory with unbroken Weyl group. Further, for higher $N$-ality representations, there are different ``$P$-orbits" of  ``$k$-strings" (both previous statements hold without accounting for screening by   heavy ``$W$ bosons").

\begin{figure}[!t] %
   \centering
   \includegraphics[width=.45\textwidth]{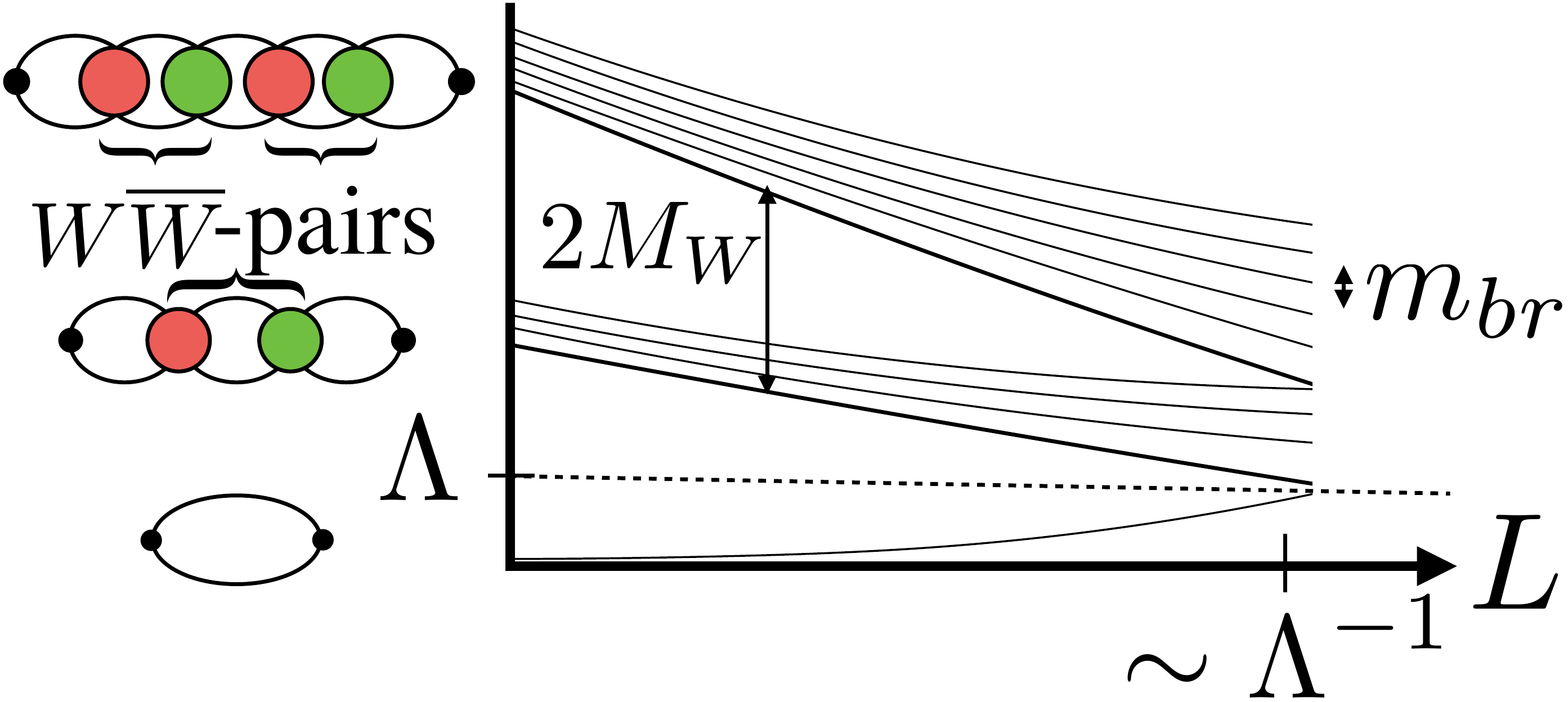} 
   \caption{A sketch of the abelian string spectrum, corresponding to the tower of $W\overline W$-bosons pairs attached to the double string, and the breather mode excitations $m_{br}$.  }
   \label{fig:string_spectrum}
\end{figure}

We leave a full taxonomy of ``$k$-strings" in QCD(adj)  for the future and briefly study  strings between fundamental quarks. From the $P$ symmetry, it suffices to take $\vec\sigma$ monodromy $2 \pi \vec{w}_1$, appropriate to the highest weight of the fundamental  (the $N_c$$-$$1$ fundamental weights $\vec{w}_k$ obey $\vec\alpha^*_p$$\cdot$$\vec{w}_k$$=$$\delta_{kp}$, $p$$=$$1$,...,$N_c$$-$$1$).
We shall argue  that these strings are also composed of  two domain walls. To this end, recall  \cite{Argyres:2012ka} that $SU(N_c)$ QCD(adj)/SYM   has  $N_c$ vacua,  $\langle\vec\sigma\rangle$$=$${2 \pi k \over N_c } \vec\rho$, $k$$=$$1,$...$N_c$, related by the broken 
$\Z_{N_c}$($\subset$$\Z_{2 N_c n_f}$) chiral symmetry. Here, $\vec\rho$$=$$\sum_{k=1}^{N_c-1}\vec{w}_k$ is   the Weyl vector and the dual photons'  periodicity   is   $\vec\sigma$$\simeq$$\vec\sigma + 2 \pi \vec{w}_k$.  An ``elementary" domain wall between the $k$-th and ($k$$+$$1$)-th vacua  then has  monodromy ${2 \pi \over N_c} \vec\rho$.
 To construct a  configuration of $2 \pi \vec{w}_1$ monodromy, we notice the identity $2 \pi \vec{w}_1$$=$${2 \pi \over N_c} \vec\rho - {2 \pi \over N_c} P \vec\rho$. A $\vec\sigma$ monodromy $2 \pi \vec{w}_1$  can now be engineered from an elementary domain wall and a $P$-transformed anti-domain wall,  as in  Fig.~\ref{fig:111}. 
A numerical minimization of (\ref{suNstring1}) confirms that, indeed, this is the string configuration in nonsupersymmetric QCD(adj) with $N_c$$=$$3,4$ (the action density plot is similar to Fig.~\ref{fig:222}).

We also note that, contrary to Seiberg-Witten theory where only linear baryons exist \cite{Hanany:1997hr}, in QCD(adj)  baryons in ``Y" or ``$\Delta$" configurations arise naturally. The affine monopole-instanton and the unbroken part of the Weyl symmetry are, again, crucial for this.  The combinatorics of such a construction   follows from the above string picture. We shall not  discuss  the energetics determining the preferred configuration here.

For $N_c$$>$$2$ SYM, the challenge is to include the now relevant $\vec\phi$-$\vec\sigma$ coupling ($\phi$ and $\sigma$ decouple only in $SU(2)$ at $g$$\ll$$1$ \cite{Anber:2014lba}); for now, we   note that   candidate string configurations with the right monodromies can  be engineered from appropriate BPS and anti-BPS walls.

\add{Similar observations to the ones in gauge theories are still true for domain walls in XY-models with the p-clock deformation which are dual to thermal gauge theories \cite{Anber:2011gn}. There, vortices would be liberated on the domain wall, which might have important consequences for the physics of transport as well as thermodynamics of these walls (e.g. heat capacity, magnetic permeability and conductivity). }

A very interesting question is how our QCD(adj)  strings   behave upon decompactification  to $\R^4$. In  SYM, no phase transition occurs and the transition to $\R^4$ should be smooth. For $n_f$$>$$1$, the $SU(n_f)$ chiral symmetry is expected to break, at least for sufficiently small $n_f$ \cite{Poppitz:2009uq} (since  fermions play crucial role in both magnetic bion formation and in stabilizing the string size, one might expect  interesting interplay between chiral symmetry breaking and confinement).  

On $\R^4$, not much is known about strings in SYM  or QCD(adj)  from field theory alone. An exception is  softly-broken Seiberg-Witten theory  \cite{Douglas:1995nw} (not pure SYM). In MQCD, the transition from softly-broken Seiberg-Witten theory to pure SYM was studied in  \cite{Hanany:1997hr}. It was found  that pure SYM strings on $\R^4$ conform, at least in the MQCD regime, to the behavior expected from nonabelian strings, with fully unbroken Weyl group  and $N$-ality-only dependent  tensions. The transition from the different abelian behaviors,  found here and in \cite{Douglas:1995nw}, to the nonabelian one should clearly involve the $W$-bosons (as they become light upon increasing $L$). Their inclusion can modify both the vacuum configurations and the confining strings themselves (a pure YM theory scenario, relating monopoles, $W$-bosons, and center vortices is in Ch.~8 of  \cite{Greensite:2011zz}).
The difficulty in pursuing this transition is, not surprisingly, the loss of theoretical control upon de-abelianization. 

It is, however, tempting to speculate, at least in SYM where continuity is guaranteed, that the gapped modes due to the double string will be responsible for the truly non-abelian structure of the string in the decompactification limit. In the abelian regime the ``non-abelian'' excitation spectrum would correspond to the exponentially small breather mode $m_{br}$, and a tower of $W$-bosons. Then, upon decompactification it is reasonable to expect the abelian string-spectrum to go into the non-abelian spectrum (see Fig. \ref{fig:string_spectrum}). \add{Needless to say, all of these questions can in principle be addressed in lattice simulations.}

{\flushleft{\bf Acknowledgements}} We thank  Soo-Jong Rey, Mithat \" Unsal and Alyosha Yung for discussions. EP acknowledges support by  NSERC and hospitality of the Perimeter Institute in the Fall of 2014. MA acknowledges the Swiss National Science Foundation. TS acknowledges the support of the DOE grant DE-SC0013036.

\bibliography{bibliography}

\end{document}